\documentclass{article}
\usepackage{graphicx}

\usepackage{amssymb}
\usepackage{amsmath}
\usepackage{latexsym}
\def\guill {\textquotedblleft ~}
\def\R{{\rm I\!R}}
\def\uc  {universal covering}
\def\ie {{i.e.}}
\def\setC{\mathbb{C}}
\def\eg {{e.g.}}
\def\andy {\mbox{~and~}}

\def\Hom {\mbox{Hom}}
\def\Harm {\mbox{Harm}}
\def\Vec {\mbox{Vec}}
\def\Harm {\mbox{Harm}}

\title{Harmonic projection and multipole  Vectors}

\author{M. Lachi{\`e}ze-Rey\\
Service d'Astrophysique, C. E. Saclay\\91191 Gif sur Yvette cedex, 
France}

\begin{document}
\maketitle
\abstract{We show that the multipole vector decomposition, recently introduced by Copi et al.,  is a consequence of Sylvester's theorem, and corresponds to the Maxwell representation. 
Analyzing  it in terms of harmonic polynomials, we  show that this decomposition     results   in fact from the application Êof the harmonic projection operator and its inverse. We derive the  coefficients of the usual harmonic decomposition from the multipole vectors. We answer to \guill an open question ", first raised  by Copi et al.,  and reported   by   Katz and Weeks, by showing that the decomposition resulting from their corollary is unstable. We propose however a new decomposition which is stable.  We generalize these results to complex functions and polynomials.}

\section{Introduction}

The recent CMB data have emphasized the necessity to handle data on the sphere,  like the temperature [fluctuations]Ê   of the CMB on the last scattering surface.  Since the role of Fourier transform on the sphere is played by the expansion  in   spherical  harmonics, this approach is widely popular. The techniques are now standard and   have been widely used for analysis and interpretation of the results. 

To analyze a function, the first step is generally a separation of scales provided by the multipole development $f=\sum _{\ell}Êf_{(\ell)}$. The sum is over all integers, but a cut-off is made at some $L$   to take into account   the resolution of the data. Ê The   spherical harmonics decomposition is completed by \begin{equation}
\label{sh}
 f_{(\ell)} \equiv \sum _{m=-\ell}^{\ell}~Êf_{\ell m}~ÊY_{\ell m}.
\end{equation}

Recent analyses (\cite{Schwarz}, \cite{Tegmark}), using various methods, have reported some signs  of anisotropy and/or non gaussianity in the CMB data. On the other hand, such effects are expected in some theoretical  models of the primordial universe. This motivates an active research for such effects in the present and future CMB data, for which the relevance of the spherical harmonics approach has been questioned.  

In addition  to the unfriendly behavior of   the  $Y_{\ell m}$  under spatial rotations,  this has  motivated     interest  towards other techniques. In this regard,  
Copi, Huterer and  Starkman  (\cite{Copi}, hereafter CHS) have recently 
shown the  association of a series of $\ell$ \guill  multipole vectors "   to    any multipole  $f_{(\ell)}$. They suggest  to represent $f_{(\ell)}$ by these vectors instead of by  the usual $f_{\ell m}$. 
Having  shown interesting properties of this set of vectors (in particular the fact that they are rotated by the usual representation), they give   convincing arguments for the relevance of their new techniques, whose interest is confirmed by application  to the CMB data.  This result  was later  confirmed and interpreted in terms of polynomials by 
Katz and Weeks (\cite{Weeks}, hereafter KW). 
However, as mentioned by both groups, many  questions remain about the signification, relevance and interpretation of this decomposition. The present paper intends to contribute to this subject.

 In section  \ref{Maxwellr}, following a recent   work by Dennis \cite{Dennis}, I  show  that the  CHS   result   is in fact a direct consequence  of  the \emph{Maxwell multipole representation}. 
 
In section (\ref{Harmonicp}), I  recall first   some well known  properties of harmonic polynomials. Then, I show that       the correspondences established by CHS and  KW   identify to the  \emph{harmonic projection} and its inverse (whose existence is established in this context). This allows a shorter proof of the decomposition of an harmonic vector, and  provides the   link  with the usual spherical harmonics approach. In particular, this allows (\ref{link}) to estimate the $f_{\ell m}$ coefficients from the multipole  vectors. 
This also allows (\ref{stab}),     to give a (negative)    answer to the \guill open question~" of  stability, raised by CHS, and   reported by   KW,  about   their multipole development; on the other hand, I    propose   a new development  which remains stable. Finally (\ref{complex}),   extension to complex polynomials is discussed. 
 
\section{ The real vector decomposition}
\subsection{Multipoles and harmonic polynomials}

Functions on the sphere $S^2$ may be seen as reductions to $S^2$ of functions on the embedding space $\R ^3$, in particular polynomials.

There are many complementary ways to consider a multipole $f_{(\ell)}$: \begin{itemize}
  \item 
as an  eigenfunction   of [the Laplacian $\Delta_{S^2}$ on] the sphere $S^2$ with   eigenvalues $\lambda _{\ell} =-\ell~(\ell+1)$:   I call $f_{(\ell)}$  an    $\ell-$eigenfunction;
  \item 
  as a function with  definite squared  angular momentum~$\lambda _{\ell}$, since $\Delta_{S^2}$ identifies with  the squared angular momentum momentum operator~$J^2$. A  further classification,  by the projection $J_z$ of the  angular momentum, leads to  the usual   $Y_{\ell m }$, as the normalized $\ell-$eigenfunction, with eigenvalue $m$ of $J_z$. 
  \item 
  as a vector of the $(2\ell +1)-$dimensional irreducible  representation of the rotation group SO(3), or of its \uc ~SU(2).
  \item 
As the reduction to $S^2$ of    an $\ell$-harmonic polynomial: an  homogeneous polynomial of degree $\ell$  (hereafter an  $\ell$-homogeneous polynomial) of $\R^3$, which is $\R^3$-harmonic, \ie,   verifies $\Delta  P=0$, where  $\Delta   $ is the Laplacian on~Ê$\R ^3$.
 
Hereafter,  I    note $\Harm (\ell)$  and $\Harm (\ell, \R)$   the  vector spaces (on $\R$ and $\setC$ respectively) of $\ell$-harmonic  polynomials with complex and real coefficients respectively.
I    note   $\Hom(\ell)$ and    $\Hom (\ell,\R)$ the  vector spaces (on $\R$ and $\setC$)  of homogeneous polynomials of degree $\ell$  (hereafter $\ell$-homogeneous polynomials ),  with complex and  real coefficients respectively. 
  \item 
As a completely symmetric traceless tensor of rank $\ell$.
 The natural development  of any $\ell-$homogeneous polynomial $P  \in \Hom(\ell)$,  \begin{equation}
\label{ }
 P(X)= \sum _{i_1,i_2,...,i_{\ell}}~ÊF_{i_1,i_2,...,i_{\ell}}~Êx^{i_1}x^{i_2} ...x^{i_{\ell}} ,
\end{equation} involves    a completely symmetric tensor  $F_{i_1,i_2,...,i_{\ell}}=F_{(i_1,i_2,...,i_{\ell})}$. Thus, any ${\ell}-$homogeneous polynomial correspond to  such a tensor. 
For an harmonic polynomial, this tensor is traceless: any ${\ell}-$harmonic  polynomial correspond to  a  completely symmetric traceless tensor. The harmonic projection (see below) is obtained by taking the traceless part. 
 \item 
An equivalence class in  $\Hom(\ell)$, see below (\ref{Harmonicp}).
\end{itemize} 

\subsection{Multipole vectors and Maxwell representation}\label{Maxwellr}

Let us consider the $\ell $-homogeneous polynomials $P$ of the form \begin{equation}
\label{vec}
P(X)=A ~ (X \cdot u_1)...(X \cdot u_{\ell}) ,
\end{equation} where $A$ is a real constant and the $u_i$ are unit [and real] vectors of $\R ^3$, \ie, points (=directions)  on the unit sphere.   
I call    \begin{equation}
\label{ } \Vec   (\ell,\R) \subset \Hom   (\ell,\R)\end{equation}  the set of such polynomials (which are in general non harmonic). 

CHS \cite{Copi}  have shown the one to one correspondence 
 \begin{eqnarray}
 \Harm   (\ell,\R) & \mapsto  &  \Vec(\ell,\R) \label{Starkman}\label{CHS} \\
H &   \mapsto & vecH ,  \nonumber \end{eqnarray}
with $vecH$ of the form (\ref{vec}) above.

They   explicit  the  correspondence  in the tensorial notation: symmetrization of the    tensor $u_{i_1}~u_{i_2}...u_{i_{\ell}}~$ gives the tensorial form of $vecH$. Then, the traceless part gives the tensorial form of $H$.  This allows to solve explicitly the correspondence (although with heavy calculations).

Following \cite{Dennis}, we show that this correspondence is a consequence of the \emph{Sylvester's theorem}. (Note that the classical proof of Sylvester's theorem implies B\'ezout's theorem, like the proof in  \cite{Weeks}. However, \cite{Dennis} provides a different proof).  The latter states that any  $\ell $-harmonic polynomial with real coefficients can be uniquely written as  \begin{equation}
\label{Sylvester}
H(X)=r^{2\ell+1}~\nabla _{u_1} \nabla _{u_2}...\nabla _{u_{\ell}}\frac{1}{r}, ~Ê\forall H \in \Harm(\ell,\R), 
\end{equation} with $r^2 \equiv X^2=X \cdot X$,  the $u_ i$ as above and the directional derivatives  $\nabla _{u_i} \equiv u_i \cdot \nabla$. This is  known as the \emph{Maxwell 
multipole representation}, and~\cite{Dennis}
has shown that it  implies the unique decomposition \begin{equation} \label{dec}
H   = vecH   +r^2 ~Q ; Q \in \Hom (\ell -2,\R).
\end{equation} 
This establishes the  correspondence (\ref{CHS}),  namely the    CHS's result.
 
 To resume,  
$$ \forall H \in \Harm(\ell,\R);$$ \begin{equation}
\label{ } H(X)=r^{2\ell+1}~\nabla _{u_1} \nabla _{u_2}...\nabla _{u_{\ell}}\frac{1}{r}=A ~ (X \cdot u_1)...(X \cdot u_{\ell}) +r^2 ~Q ; 
\end{equation} $$Q \in \Hom (\ell -2,\R).$$

The proposition by CHS,  and then by KW, to characterize any $H$ (or, equivalently, $f_{(\ell)}$)  by  the constant $A$ and the $2\ell$ components of the $u_i$ (rather than the $2\ell+1$ $a_{\ell m}$'s)  offers the 
 main  advantage  that the $u_i$ rotate  under the $\R^3$ representation (which corresponds to $\ell=1$). As shown explicitly  by \cite{Dennis}, this corresponds to a factorization of the [spin]  $\ell$ representation into a product of  $\ell=1$ representations. 
 
 Now we show now that this decomposition corresponds to  the  harmonic projection.
 
\subsection{Harmonic projection}\label{Harmonicp}
  
The  well known decomposition (see, \eg, \cite{vilenkin})  \begin{equation}
\label{decomposition}
\Hom(\ell) =\Harm ({\ell}) \oplus r^2 ~Ê\Hom(\ell-2) \end{equation}  means that any $\ell$-homogenous   polynomial $P \in \Hom (\ell)$ can be uniquely decomposed as  \begin{equation}
\label{Harmonic projection}
P=\Pi P+r^2~Q;~Ê\Pi P \in \Harm(\ell),~Q \in \Hom (\ell -2).
\end{equation}  The   $\ell$-  harmonic $\Pi   P$ is called the \emph{harmonic projection} of $P$       \cite{vilenkin}, and $Q $ is $(\ell-2)$-homogenous.    It can be checked that $\Pi  :  \Hom(\ell) \mapsto \Harm(\ell)$ is effectively a projection operator (which is not inversible). 

{\bf An equivalence relation}

We may define an    equivalence relation in $ \Hom(\ell)$:  ,\begin{equation}
\label{ } P \approx P' \Leftrightarrow  P-P' = r^2~ÊQ; ~ÊQ  \in  \Hom(\ell-2). \end{equation}
Two   $\ell$ - homogeneous polynomials  are equivalent if their difference is a multiple of the monomial 
$r^2 \equiv X \cdot X$. 
Then it is easy to check that two polynomials are equivalent iff they have the same  harmonic projection. Thus the vector space $\Harm ({\ell}) $ appears as the quotient of $ \Hom ({\ell}) $ through the equivalence relation: each $\ell$-harmonic polynomial represents an equivalence class in $ \Hom ({\ell})$.
  
In tensorial notations,   the harmonic projector  $\Pi$  takes the traceless part of the tensor (see, \eg,  \cite{vilenkin}).    This allows to infer     \begin{equation}
\label{ } H = \Pi (vecH).\end{equation}

All these relations are similarly verified by polynomials with real coefficients.

The correspondence being  one to one (Sylvester's theorem), we may  invert   [the reduction to $\Vec(\ell,\R)$ of] the  $\Pi$ operator.  This provides the    interpretation of  CHS's result as the action of  $\Pi   ^{-1}$: 
 \begin{eqnarray}
 & \Harm   (\ell,\R) &\leftrightarrow    \Vec(\ell,\R)  \nonumber \\
\Pi:& H \equiv \Pi vecH& \leftarrow     vecH ,  \nonumber\\
\Pi^{-1}:&~ÊH & \mapsto   vecH \equiv \Pi^{-1}H . \label{Starkman2} \end{eqnarray}

This allows   to see also $\Vec(\ell,\R)$ as the set of  equivalence classes in      $\Hom (\ell,\R)$.
Note however  that 
$\Vec(\ell,\R)$ is not a subvector space of $\Hom(\ell,\R)$. But we may  provide   a vector space structure to $\Vec (\ell,\R)$ by defining the \guill sum " :\begin{equation} \label{ }
P ~\tilde{+}  ~Q \equiv \Pi   \left[  \Pi   ^{-1} P+ \Pi  
 ^{-1} Q \right], ~ÊP,Q \in \Vec  (\ell,\R). \end{equation}
For instance, we have $z^2~ \tilde{+} ~y^2=-x^2$.

\subsection{Link with the multipole coefficients}\label{link}

The CHS's formula allows to characterize any multipole $f_{(\ell)}$ by  the   constant $A$ and the coordinates of the $\ell$ unit vectors.  It is interesting to make the link with the other description given by the $2 \ell +1$ coefficients $f_{\ell m}Ê$ of the usual harmonic decomposition (\ref{sh}).
 
  We start from  the identity
\begin{equation}
\label{ }  
 a_1  ~Ê
a_2 ~Ê...a_\ell =\frac{1}{B}~Ê \sum_{\epsilon _1~\epsilon _1~...\epsilon _{\ell}}  (\epsilon _1~Êa_1 ~Ê
+\epsilon _2~a_2 +Ê...+\epsilon _{\ell}~a_{\ell}) ^{\ell};~ÊB=_{def}~Ê\ell!Ê\sum_{\epsilon _1~\epsilon _1~...\epsilon _{\ell}} 1,\end{equation}    where each $\epsilon_{i}$ takes the two values  -1 and 1. It implies\begin{equation}
\label{ }  
(u_1 \cdot X)~Ê
(u_2 \cdot X)~Ê...(u_\ell \cdot X)=\frac{1}{B}~ \sum_{\epsilon _1~\epsilon _1~...\epsilon _{\ell}} [(\epsilon _1~Êu_1 ~Ê
+\epsilon _2~u_2 +Ê...+\epsilon _{\ell}~u_{\ell}) \cdot X]^{\ell}.\end{equation}  Assuming unit vectors, this may be rewritten 
$$\frac{1}{B}~  \sum_{\epsilon _1~\epsilon _1~...\epsilon _{\ell}} ~Ê(\epsilon _1~Ê\cos \theta_1 ~Ê
+\epsilon _2~\cos \theta_2 +Ê...+\epsilon _{\ell}~\cos \theta_{\ell})   ^{\ell},
$$ with $\cos \theta _i =u_i \cdot X$. From the calculation in the appendix, it results that 
$$\Pi [  (u_1 \cdot X)~Ê
(u_2 \cdot X)~Ê...(u_\ell \cdot X)]=  \sum _m a_{\ell m} ~ÊY_{\ell m}  (X),~$$ with \begin{equation}
\label{ }
a_{\ell m}  = \frac{K_{\ell}}{B}~
\sum_{\epsilon _1~\epsilon _1~...\epsilon _{\ell}}  Y_{\ell m}^* (\epsilon _1~Êu_1 ~Ê
+\epsilon _2~u_2 +Ê...+\epsilon _{\ell}~u_{\ell})  ,
\end{equation} with $K_{\ell}$ given by (\ref{kl}).

This formula gives the coefficients of the harmonic decomposition  as a function of the  multipole vectors. The reciprocal calculation can be made in the tensorial formalism as indicated by CHS.

 \section{Extension to homogeneous Polynomials}
 
By (\ref{decomposition}), an arbitrary   $\ell-$homogeneous polynomial  $P \in \Hom  (\ell,\R)$ is uniquely projected to 
 $\Pi P  \in \Harm  (\ell,\R)$, with $ \Pi P =P + r^2~ÊQ'$,  $Q' \in \Hom  (\ell-2,\R)$. On the other hand, CHS result implies \begin{equation}
\label{ }
\Pi P=   vec (\Pi P)+ r^2~Q'',~vec( \Pi P) \in  \Vec  (\ell,\R),
~ÊQ'' \in \Hom  (\ell-2,\R).\end{equation} These two relation imply \begin{equation}
\label{ }
P = \Pi ^{-1} \Pi P+r^2~ÊQ;~Q \in \Hom  (\ell-2,\R). \end{equation}

This extension of  CHS's result to $\Hom (\ell,\R)$ had  been found by  KW. The present derivation   provides a   shorter demonstration:    any real $\ell-$ homogeneous polynomial  $P \in \Hom  (\ell,\R)$ has the   unique decomposition    \begin{equation} \label{Weeks}
P  =vecP   + r^2~ÊQ,
\end{equation} with    $vecP   \in \Vec  (\ell,\R)$  and $Q \in \Hom  (\ell -2,\R)$.  The CHS's result appears as a special case.

 \subsection{ Stability }
\label{stab}

For any  function $f$ on $S^2$, like for instance the temperature of the CMB, let us call \begin{equation}
\label{Mexp}f_{\le L} \equiv \sum_{\ell=0}^L f_{(\ell)}
\end{equation} its approximation by the $L$ first multipoles. 
For each value of $L$, $f_{\le L}$ may be developed as a sum of vectorial polynomials, according to the Corrolary  2 of KW.  Do the vectorial polynomials in the development change when $L$ increases ? This is the stability problem asked by CHS and KW  (their section VI: \guill an open question "), to which we answer here.

To answer the question, let us consider two successive approximations  $f_{\le L}$ and $f_{\le L+1}$ of the same function $f$. According to the KW's Corollary~2, they can be developed uniquely as $$f_{\le L}=\sum _{\ell =0}^L~V_{\ell} 
\mbox{~and~} f_{\le L+1}=\sum _{\ell =0}^{L+1}~W_{\ell},$$ where $V_{\ell}, W_{\ell} \in \Vec (\ell)$.
As quoted by KW, $W_{L+1} \ne V_{L+1}=0$, $W_{L} = V_{L}$ and $W_{L-1} \ne V_{L-1}$ in general. The question of stability concerns the possible equality between the other  $V_{\ell}$ and $W_{\ell}$,  for $\ell < L-1$.

Let us assume stability, \ie,  that  \begin{equation}
\label{condition}V_{\ell}=W_{\ell}, ~\forall \ell <L-1 .\end{equation} This  would  imply \begin{equation}
\label{rrt}
 f_{( L+1)}= f_{\le L+1}- f_{\le L }= W_{L+1}+W_{L-1}-V_{L-1}.\end{equation}

First, let us note that, for a given  ÊÊ$f_{\le L}$ fixed,  $ f_{( L+1)} $  can be [the reduction to $S^2$ of] any $(L+1)$-harmonic polynomial. On the other hand, the relation (\ref{rrt}) would hold for any $f_{\le L }$, and in particular when $V_{L-1}=0$. This would imply that   $f_{( L+1)}=  W_{L+1}+W_{L-1}$ would hold for any $(L+1)$-harmonic polynomial. In other words, \begin{equation}
\label{ }
\forall H \in \Harm(L+1), ~ÊH= V_1+V_2,~V_1 \in \Vec(L+1),~V_2 \in \Vec(L-1):\end{equation} the   decomposition, which is unique, would involve two terms only. This  is clearly wrong and, thus, the stability hypothesis (\ref{condition}) is not true: no term in the decomposition is stable (excepted in the very special case, where all the monomial on the decomposition of $P$ are already in vectorial form). 

{\bf Expansion of  the exponential}

To judge the severity of this instability, it may be convenient to examine a simple example, namely the case of the exponential $f(x) \equiv \exp (k.x)$, where $k$ and $x$ are vectors of $\R^3$. The usual development of the exponential gives in fact its exact multipole vector expansion (in this case, as expected, the unit vector $\hat{k} \equiv k/\mid k\mid =k/K$  is the unique one appearing):\begin{equation}
\label{true}
f(x)=\sum _n ~Ê\frac{K^n}{n!}~Ê(\hat{k}\cdot x)^n.
\end{equation}

On the other hand,   the well known  multipole decomposition of the exponential implies (on the sphere) \begin{equation} \label{}
f_{(\ell)}(x)=(2\ell +1) ~Êi^{\ell}~j_{\ell}(K)~ÊP_{\ell}(-i\hat{k}\cdot x).
\end{equation}
Here, $j_{\ell}$ is the spherical Bessel function, and the Legendre Polynomials $P_{\ell}$ are extended  to complex arguments, insuring that  $f_{(\ell)}(x)$ takes real values. It results 
\begin{equation} \label{ }
f_{\le L}(x)=~\sum_{\ell=0}^L~(2\ell +1)~Êi^{\ell}~j_{\ell}(K)~ÊP_{\ell}(-i\hat{k}\cdot x).
\end{equation}
The usual expansion of the Legendre Polynomial allows  to estimate the higher ($L$) order term in this development, as  
\begin{equation}
\label{approx}
(2L +1)~j_{L}(K)~\frac{(2L)!}{2^L~Ê(L!)^2}Ê(\hat{k}\cdot x)^L
\end{equation}
(the imaginers disappear). 

\begin{figure}[tb]
\includegraphics[width=5in]{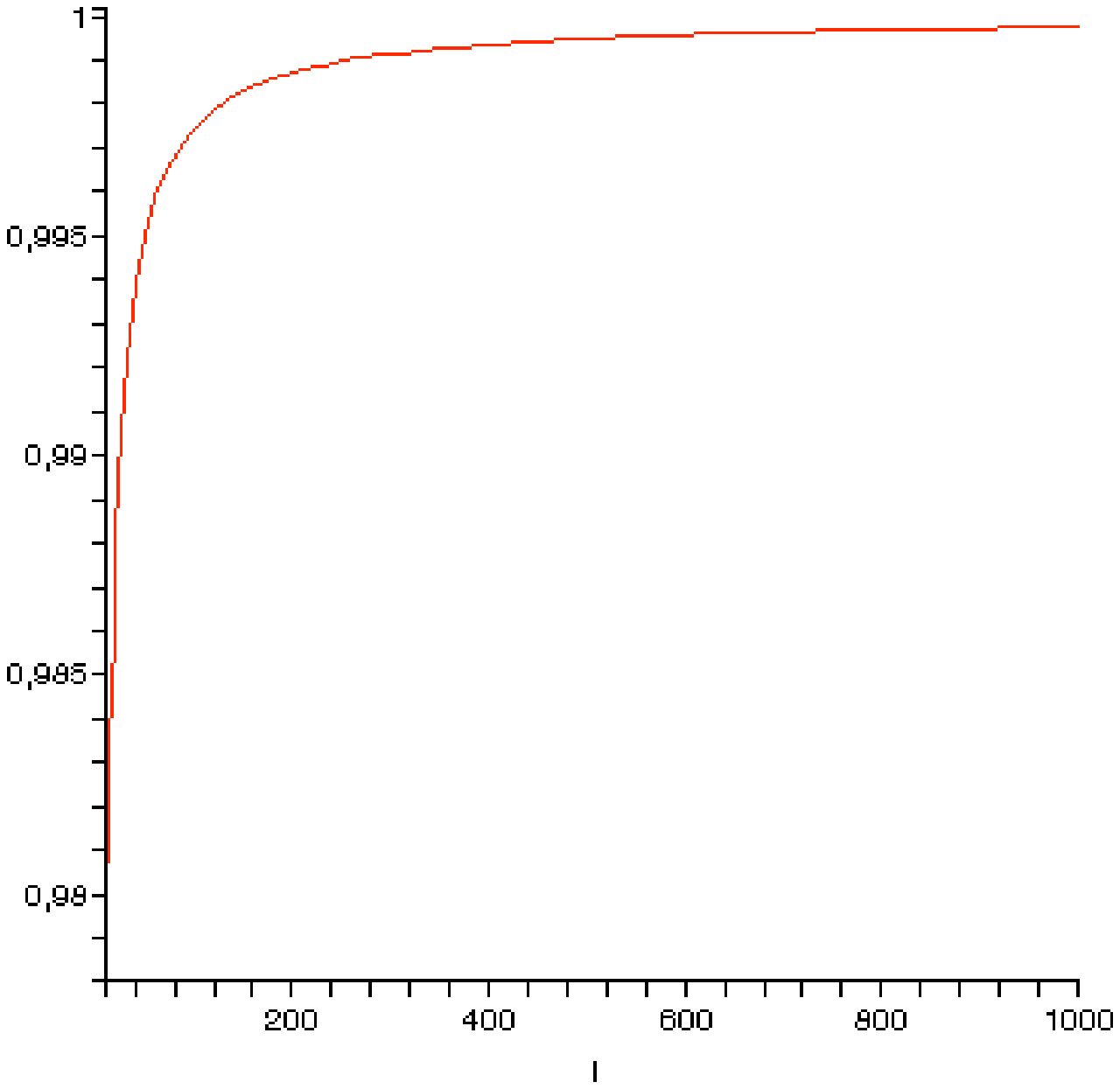}
\caption{The ratio $R$ as a function of the multipole index $\ell$, for $K=1$}
\label{figure1}
\end{figure}

This is the leading term in the decomposition of  $f_{\le L}$ according to the corollary of KW. Thus, in this simple case, we may compare the term of order $L$ in the expansion  of  the \guill complete " function $f$, given by (\ref{true}), with the  corresponding term  (\ref{approx}) in the order $L$ approximation $f_{\le L}$. Their ratio is given by 
$$R \equiv    \frac{j_{L}(K)}{K^L}~\frac{(2L+1)!}{2^L~Ê(L!) }.$$ It differs from 1,  but it tends towards unity when $\ell$ goes to infinity, as shown in   Figure \ref{figure1} (in the case $K=1$). 

\subsection{A stable decomposition}

This instability motivates the search for  an other decomposition, which is in fact provided by the Maxwell representation . 
 Starting from the multipole expansion (\ref{Mexp}) of a function $f$, we simply replace each element by its Maxwell representation. This gives the   decomposition
$$f_{\le L}(x)=\left[ \lambda _0~Êr~Ê+~Ê \lambda _1~Êr^3~Ê\nabla _{u_{1,1}}+...+
\lambda _{\ell}~Êr^{2\ell+1}~Ê\nabla _{u_{\ell ~,1}}...\nabla _{u_{\ell ~,\ell}}+... \right.$$ \begin{equation}
\label{ } \left.
...+
\lambda _{L}~Êr^{2L+1}~Ê\nabla _{u_{L ~,1}}...\nabla _{u_{L ~,L}} \right] ~Ê(\frac{1}{r}), \end{equation}
which  is stable by construction. (I thank Jeff Weeks for the idea of this concise  demonstration).

\subsection{Anisotropy}

The multipole vectors offer an immediate advantage: they are rotated by SO(3) as  vectors, \ie, according to the usual vector representation. Thus, they seem very appropriate to study the possible anisotropy of the CMB data. 

For instance, when $T$ is a Gaussian isotropic  random field, all the  $T_{(\ell)}$ are independent functions. Thus no correlation should exist between vectors corresponding to different multipole values $\ell$. The situation is not so simple for the different vectors in the decomposition of a given multipole.  In the absence of any   preferred direction, no peculiar vector could appear  in the decomposition. Thus, for any realization, the vectors obtained should be isotropically distributed. For the quadrupole,  \cite{mag}  have studied the algebraic and statistical independences of the coefficients in this decomposition, for a Gaussian random process. 

This  gives a high significance to  the results that  CHS and KW report, from  their analyses,    of an \guill astonishing~Ê"
quadrupole and octopole alignment. This   clear breaking of isotropy   sets the question to interpret it as a chance effect, a contamination of the data, or a cosmological effect. In this latter regard, it is tempting to invoke  an universe with multi-connected spatial topology, where the large scale isotropy disappears, and is replaced by partial isotropy, \ie, symmetry under a group $H$ related to the  holonomy group  $\Gamma$.  It is however not so clear that this may explain these results. 

It is well known that a 
 multi-connected space imposes a quantification of the wave vectors appearing in the mode decomposition of spatial fluctuations, which is reflected in the distribution of angular temperature fluctuations of the CMB. The statistical distribution of the latter must therefore be invariant under~$H$. This has for consequence that, for a given multipole, the distribution of the vector $u_i$ must be $H$-invariant, with the immediate consequence that the \emph{number}Ê of such vectors must be an integer divisor of the order $n_{H}$ of $H$. This implies that, for a perfectly representative distribution, the vectors $u_i$ can be present only if $\ell$ (or a divisor of $\ell$) divides $n_{H}$. For instance, if the  quadrupole and octopole alignment is significative in this regard, this would imply that   $n_{H}$ is a multiple of 6, which puts strong constraints on $H$, and thus on $\Gamma$. 

 The question of anisotropy has also been treated  by 
\cite{mag}. They propose to extract from the set of vectors $u_i$ (and the constant $A$) related to a given value of $\ell$, a set of $2\ell -2$ rotation invariant quantities plus 3  rotational degrees of freedom.  This interesting suggestion, however, only provides a partial answer to the question of anisotropy. Firstly, the rotation invariant quantities provide information on anisotropy.  For   example, a multi-connected space   will impose some fixed  angles 
between the vectors, in order that the set of vectors be $H$ invariant (for instance, right angles in a toroidal universe). Thus, even if they are rotation invariant,  the 
scalar products between the vectors  bring  valuable information about anisotropy. On the other hand, the examination of (for instance) the rotation properties of \guill anchor vectors " when $\ell$ increases, has no simple interpretation. For instance, in the simple case of a toroidal universe, the different multipoles (different $\ell$'s) may favor vectors which are the sides, or the diagonals, or different preferred vectors.  In this case, despite a strong and well defined anisotropy, the anchor vectors would show no alignment (although they would exhibit some specific and  well defined correlations which, however, cannot only be  predicted a priori with a  specified   particular model).
Thus we conclude that the interpretation of the multipole vectors in terms of isotropy/anisotropy of the data remains an open question. 

In fact (excepted for $\ell=1,2$), there is no standard way  to associate to the  $\ell$ multipole a  definite spatial   orientation (which belongs to the  3  dimensional  representation of SO(3)). The only  exceptions are  the dipole ($\ell=1$), which selects   one preferred direction;  and the quadrupole, which selects  a preferred orthonormal frame (see also \cite{Dennis} for an interpretation of the dipole and quadrupole  vectors).
 In fact, an $\ell$  multipole  may be seen  \cite{lach} as a function on the  fuzzy (non commutative) sphere $S_{fuzzy,2\ell+1}$, which is itself  an approximation of the ordinary  sphere by a set of $2 \ell +1$ cells.
 
\section{ Extension to complex polynomials}\label{complex}

The results of CHS and KW were obtained  for   polynomials, and  vectors in the decomposition, with  \emph{real} coefficients. 
Since, for instance, the   usual spherical harmonics correspond to  polynomials with complex  (not real) coefficients, it seems interesting to try to generalize the previous results.  (Note that  a basis of real spherical harmonics offers a limited interest, because they  are  not  eigenvectors   of the projected angular momentum).

The KW' s demonstration is based on the existence of $2 \ell $ common roots (in $\setC P^2$), to the two equations  $P(X)=0 $ and   $X^2=0$. This remains true when  the coefficients of $P(X) $ are   complex. However, in this case, the roots are no more  two by two complex conjugates. Although a similar   demonstration can be performed, there are  two main differences : 

- it remains true that the $2 \ell$ roots can be grouped   by pairs, to generate $\ell$ lines. But there is no canonical way to do it, as in the real case (associating  the complex conjugates). The consequence is that the resulting  decomposition is  not  unique:  there are  $N$ different decompositions, $N$ being the number of distinct ways to group the $2 \ell$ roots (which may be distinct or not) in pairs.\\ 
-  In general, the   roots are   not complex conjugates. Thus the   constructed  lines   are  not real. This means that the vectors obtained in the decomposition have complex coefficients. 

We obtain thus the following result : 

An $\ell$-homogeneous polynomial with complex coefficients, $P \in \Hom(\ell)$ can be decomposed in a finite number of different ways, under the form 
 \begin{equation}
\label{Weeks}
P(X) =    \alpha~Ê(X \cdot v_1)...(X \cdot v_{\ell}) + X^2~ÊQ,
\end{equation} where  the $v_i$  have now complex coefficients and $Q \in \Hom(\ell -2)$. Given the normalization constant $\alpha \in \setC$, it is not a restriction to assume these vectors unitary ($v^2=1$)  or null ($v^2=0$). 

When $P$  has real coefficients, one   decomposition (among the others) is canonical, and  involves only real (unit) vectors : this is the  KW's result. There are however, in general,  other decomposition involving complex vectors, two by two conjugates. 

\subsection{Some examples}\begin{itemize}
  \item 
$P=x^2+y^2$ has the real decomposition $ P=-z^2+r^2$, but also the complex one $P=(x+i y)(x-i y)$.\\
\item 
$P=x^2-y^2$ has the real decomposition $P=(x+y)(x-y)$ , but also the complex one $P=-(2y^2+z^2) +r^2= -(\sqrt{2}~Ê y- i  z )(\sqrt{2}~Ê y+i  z )+r^2$.
\item 
For  the toy quadrupole of KW, their equation (19), there are, as expected,  the 4 roots  given by their equation (22).  By associating  the complex conjugates,  they obtained  their  decomposition (26) with real vectors. But different associations provide  the two other decompositions:  $$1/6 [x(1-i\sqrt{2}) +3~y + z~(1-i \sqrt{2})]~Ê[x(1+i\sqrt{2}) +3~y + z~(1+i \sqrt{2})]+3 r^2/2$$ and 
 $$-1/2 [-x(1+i\sqrt{2}) + y + z~(-1+i \sqrt{2})]~Ê[x(-1+i\sqrt{2}) + y - z~(1+i \sqrt{2})]+ r^2/2.$$
  \item 
The toy octopole of KW  (27) gives the  6 roots  (29,30).  The association of the  complex conjugates gives   the  decomposition (32) with real vectors.  The different groupings give
$$(i~x+y-z)~(-i~x+y-z)~/2+r^2~(y-z/2),$$and$$
 -(i~x+y+z)~(-i~x+y+z)~/2+r^2~(y+z/2).$$ 
\end{itemize}

\subsection{The spherical harmonics}
 
Note that the operators  $J_i \equiv -i \epsilon_{ijk}~x^j~\partial _k$, and thus $J^2$, commute with $\Delta$  and with $\Pi$. From this, it results that the spherical harmonics
are given by \begin{equation} \label{ }
Y_{\ell m }=\Pi ~Êz^{\ell -m} ~Ê(x+i y)^m.
\end{equation}In particular,\\
$Y_{\ell \ell }=\Pi ~Ê  ~Ê(x+i y)^{\ell}=(x+i y)^{\ell}, $\\
$Y_{\ell \ell -1 }=\Pi ~Êz  ~Ê(x+i y)^{\ell-1}=z  ~Ê(x+i y)^{\ell-1}, $\\
$Y_{\ell \ell -2 }=\Pi ~Êz^2  ~Ê(x+i y)^{\ell-2}=z^2  ~Ê(x+i y)^{\ell-1}-r^2/3  ~Ê(x+i y)^{\ell-2}, $\\
...\\
$Y_{\ell 0 }=\Pi ~z^{\ell}= \frac{ Ê2^\ell~(\ell!)^2}{ Ê(2\ell!)}
P_{\ell}(z), $\\
 
 \section{Conclusion}

As proven by previous results, the multipole vector decomposition offers a new and promising technique. However, the instability of the    decomposition, and the difficulty to interpret anisotropy, 
clearly demand deeper analysis. In this regard, the result presented here, namely the link with the Maxwell representation, the interpretation in terms of harmonic projection, the established correspondence with the usual harmonic development give some new perspectives. 

It remains to analyze the practical relevance of the new stable decomposition proposed here; also, the link between anisotropy and multipole vectors deserves further exploration, in particular in the frame of cosmological models with non trivial topology.

Moreover, the scalar product  $X \cdot u$, as a function of $X$, shows some similarities with a wavelet on the sphere. Thus, the multipole vector decomposition shows analogies with   a product of wavelets. Since there is presently   a lot of interest toward wavelets on the sphere, a further exploration of these analogies may appear fruitful.  The construction of a new formalism using the multipole vector decomposition  appears therefor as a very promising task.

\section{Appendix}

By definition, $\Pi(u\cdot X) ^{\ell}$,  is $\ell -$harmonic and can  thus be expanded in spherical harmonics as $\sum _m a_{\ell m}(V)~Y_{\ell m}(X)$, for $u$ and $X$  on the sphere.
Using spherical coordinates, we have $$X = (\cos \theta,~Ê\sin \theta~Ê\cos \varphi,\sin \theta~Ê\sin  \varphi)\andy 
u = (\cos \theta ',~Ê\sin \theta '~Ê\cos \varphi ',\sin \theta '~Ê\sin  \varphi ').$$  Thus, $\Pi  (X\cdot u) ^{\ell}= 
\Pi[ \cos \theta~Ê\cos \theta '+Ê\sin \theta~Ê\sin \theta '~Ê\cos (\varphi-\varphi ') ]$ depends  on  
$ \varphi '$ only through $ \varphi-\varphi ' $.  Using $Y_{\ell m}(X)=K_{\ell m} ~e^{ im \varphi}~ÊÊP_{\ell m}(\cos \theta)$, where the $K_{\ell m}$'s  are  the usual normalization constants and the $P_{\ell m}$'s  the Legendre functions, this implies  
$$\Pi (X\cdot u) ^{\ell}= \sum _m F_{\ell m}(\theta ')~e^{-im\varphi '}~ÊY_{\ell m}(X)= \sum _m F_{\ell m}(\theta ')~e^{ im(\varphi-\varphi ')}~ÊK_{\ell m} ~ÊP_{\ell m}(\cos \theta) .$$  By symmetry, we have also 
$$\Pi  (X\cdot u) ^{\ell}= \sum _m F_{\ell m}(\theta)~e^{ im(-\varphi+\varphi ')}~ÊK_{\ell m} ~ÊP_{\ell m}(\cos \theta ') .$$ This implies, using symmetry arguments, 
$$\Pi (X\cdot u) ^{\ell}= K_{\ell}~\sum _m Y^*_{\ell m  }(u)~ÊY _{\ell m  }(X).$$
The normalization constant \begin{equation}
\label{kl}
K_{\ell}=\frac{4 \pi~Ê2^\ell~(\ell!)^2}{(2\ell+1)~Ê(2\ell!)}
\end{equation} can be calculated  from (\ref{dec}), using the polynomial development of $P_{\ell  }$.

{}

\end{document}